\documentclass[%
 reprint,
superscriptaddress,
 amsmath,amssymb,
 aps,
prl,
]{revtex4-1}

\usepackage{graphicx}
\usepackage{dcolumn}
\usepackage{bm}

\begin{document}

\title{Charge carrier inversion in a doped thin film organic semiconductor island}

\author{Zeno Schumacher}
 \email{zenos@physics.mcgill.ca}
 \altaffiliation[Present address: ]
{Department of Physics, Institute of Quantum Electronics, ETH Zurich, Switzerland}
\affiliation{%
 Department of Physics, McGill University, Montreal, Canada
}%
\author{Rasa Rejali}%
\affiliation{%
 Department of Physics, McGill University, Montreal, Canada
}%
\author{Megan Cowie}%
\affiliation{%
 Department of Physics, McGill University, Montreal, Canada
}%
\author{Andreas Spielhofer}%
\affiliation{%
 Department of Physics, McGill University, Montreal, Canada
}%
\author{Yoichi Miyahara}%
\altaffiliation[Present address: ]
{Department of Physics, Texas State University, San Marcos, Texas}
\affiliation{%
 Department of Physics, McGill University, Montreal, Canada
}

\author{Peter Grutter}%
\affiliation{%
 Department of Physics, McGill University, Montreal, Canada
}%

\date{\today}

\begin{abstract}
Inducing an inversion layer in organic semiconductors is a highly nontrivial, but critical, achievement for producing organic field-effect transistor (OFET) devices, which rely on the generation of inversion, accumulation, and depletion regimes for successful operation. In recent years, a major milestone was reached: an OFET was made to successfully operate in the inversion-mode for the first time \cite{Lussem2013}. Here, we develop a pulsed bias technique to characterize the dopant type of any organic material system, without prior knowledge or characterization of the material in question. We use this technique on a pentacene/PTCDI heterostructure and thus deduce that pentacene is n-doped by impurities. Additionally, through tip-induced band-bending, we generate inversion, depletion, and accumulation regimes over a 20~nm radius, three monolayer thick n-doped pentacene island. Our findings demonstrate that nanometer-scale lateral extent and thickness are sufficient for an OFET device to operate in the inversion regime.
\end{abstract}

\maketitle

Currently, of the few tools available to investigate charge transport and carrier generation at the nanoscale, the most versatile and powerful is the combination of non-contact atomic force microscopy (nc-AFM) and Kelvin probe force spectroscopy (KPFS): the  former provides structural information, while the latter allows for measuring charge distribution. KPFS is a measurement of the contact potential difference between the AFM tip and the sample, and so, can be used to detect surface potential changes that occur when a semiconductor is illuminated \cite{Takihara2008,Borowik2014,Loppacher2005,Shao2014,Hoppe2005,Hoppe2006,Neff2015,Luria2012a}. The combination of these techniques opens the door to understanding the fundamental physics and properties of nanoscale electronics, and has been used to measure the charge state of individual molecules \cite{Steurer2015}.

Nc-AFM can be used to directly measure the force gradient between the tip and sample. In the small oscillation amplitude limit for the familiar metallic tip-metallic sample case, the electrostatic force, $F_{elec}$, can be related to the measured frequency shift, $\Delta f$, as follows

\begin{equation}
\label{eq:Felec}
\Delta f \propto \frac{\partial F_{elec}}{\partial z} = \frac{1}{2} \frac{\partial^2 C_{ts}}{\partial z^2} (V_{DC} - V_{CPD})^2,
\end{equation}
where $C_{ts}$ is the tip-sample capacitance, $V_{DC}$ the applied bias voltage, and $V_{CPD}$ the contact potential difference. Usually, the tip-sample capacitance is assumed to be independent of the applied voltage---and for many doped semiconductor systems studied by nc-AFM, this is a good approximation, even though the capacitance is strictly a function of the applied voltage in these cases \cite{Sadewasser2009,Hudlet1995}. 

A more rigorous treatment considers that the capacitance of a doped semiconductor varies as the applied voltage is increased. In a doped semiconductor, band bending at the surface and the formation of depletion and inversion layers within the penetration-depth lead to this voltage-dependent capacitance. 

There is clear dissonance between the above-stated physical picture and the common analytical assumption that the tip-sample capacitance is voltage-independent\cite{Sadewasser2018}.  The solution lies in noting that the effects of a voltage-dependent capacitance are, in most systems, quite  benign. The tip-induced band bending in the semiconductor needs to be strong enough to achieve depletion or even inversion. The tip-induced band bending is a function of applied bias voltage, tip-sample distance, tip radius, as well as semiconductor parameters such as doping concentration \cite{Feenstra2006,Feenstra2003}. Therefore, only for the right set of these parameters depletion can be achieved, resulting in a change in the tip-sample capacitance. Furthermore,  it is necessary to record the full $V_{DC}$ bias response of the system to observe the effect of the voltage dependent capacitance. As such, the popular Kelvin probe force microscopy (KPFM) technique is not appropriate for this application, as it only records the contact potential difference between the tip and the sample (i.e. the maximum of the parabola and not the curvature).

The first foray in understanding the fundamental physics of doped semiconductors probed by AFM came in 1992 \cite{Huang1992b,Hudlet1995,Donolato1996a}. Experimental verification of the theorized behavior was published several years later, when the accumulation, depletion, and inversion regimes in a n-doped InAs sample were measured using AFM \cite{Schwarz2000}. To date, no such effect has been measured in organic semiconductors with AFM.

\begin{figure*}
\centering
 \includegraphics{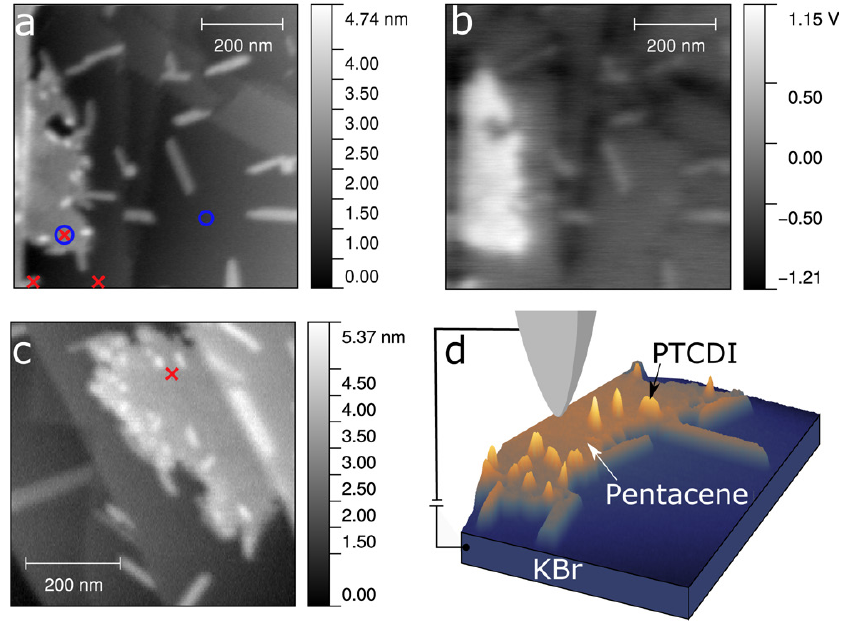}
\caption{a) Topography image of PTCDI on a pentacene island, on top of a KBr substrate 
(oscillation amplitude, $A = 6~\text{nm}$, frequency shift setpoint, $\Delta f = -6~\text{Hz}$). 
b) Simultaneously acquired frequency modulation KPFM image 
(frequency and amplitude of the ac modulation: $600~\text{Hz}$ and $1~\text{V{pp}}$). 
c) Topography image of an additional measurement site ($A = 6$\,nm, $\Delta f= -10$\,Hz). 
Red crosses and blue circles indicate locations of spectroscopy or pulsed measurements, respectively. 
d) Graphical illustration of the measurement setup.}
\label{fig:topo}
\end{figure*}

In this study, we thermally deposit 0.2 monolayer of pentacene on an in-situ cleaved KBr crystal, 
followed by evaporation of 0.4 monolayer of 3,4,9,10-perylenetetracarboxylic diimide (PTCDI). In 
general, pentacene grows in large, flat islands on KBr and stands upright, with the plane and the long 
axis of the molecule perpendicular to the substrate surface, or with the plane of the molecule 
perpendicular to slightly tilted and the long axis parallel to the surface \cite{Neff2014}. 
Thus, for a measured height of approximately 2.1~nm, we expect the pentacene islands are three monolayers thick, 
assuming the molecular long axis is parallel and the plane of the molecule perpendicular to the surface.
PTCDI, on the other hand, grows on KBr substrates to form longer, needle-shaped islands 
\cite{Topple2011}. As such, pentacene and PTCDI are easily distinguishable in topographic images. Together pentacene and PTCDI form heterostructures of small PTCDI islands at the edges or on top of large pentacene islands. Pentacene together with PTCDI derivatives have previously been used as model structures for ambipolar OFET \cite{rost_ambipolar_2004}. A
topography image, recorded in constant frequency shift 
mode, is shown in Figure~\ref{fig:topo}a; simultaneously acquired frequency modulation(FM)-KPFM data is shown in
Figure~\ref{fig:topo}b.

\begin{figure*}[htp]
\centering
\includegraphics[width=1\linewidth]{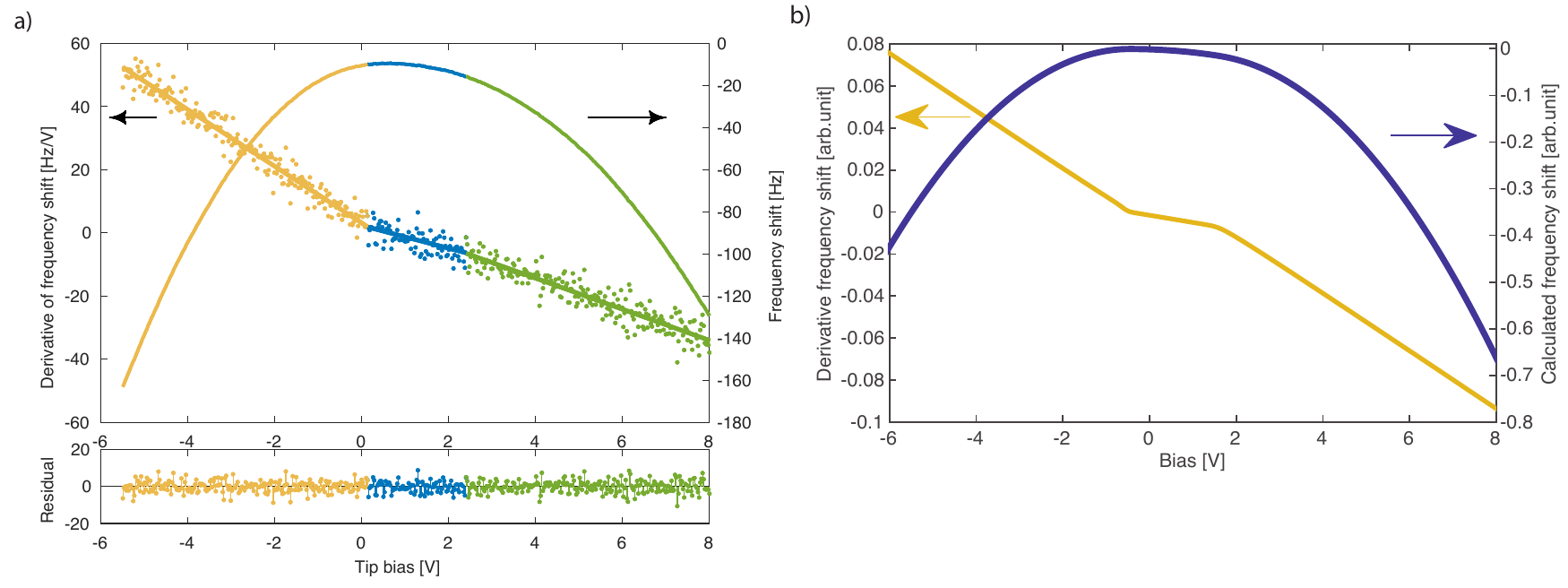}
\caption{a) KPFS measurement on a pentacene/PTCDI heterostructure. A clearly non-parabolic response is observed. The numerically differentiated frequency shift with respect to the applied bias is fitted to a straight line from the positive(green)/negative(yellow) side until the fit is optimized. The center (blue) is fitted with a straight line between the end points of the respective fits. The residuals are shown below. b) Calculated frequency shift for a doped semiconductor-metal tip system based on \cite{Hudlet1995}. An intrinsic carrier density of $10^{2} (\text{cm}^{-3})$, a dopant density of $(5 \times 10^{15}\text{cm}^{-3})$, a mean tip-sample distance of 5.5~nm and an oscillation amplitude of 9~nm was used.}
\label{fig:asym_bias}
\end{figure*}

We perform KPFS on a pentacene/PTCDI heterostructure (the location is indicated in
Figure~\ref{fig:topo}c by a red cross) as shown in Figure~\ref{fig:asym_bias} a) (recorded at 0~nm tip lift from $\Delta f = -10~\text{Hz}$ setpoint). 
The frequency shift is not parabolic as a function of applied bias 
as expected from equation \ref{eq:Felec}, indicating a non-negligible voltage-dependence in the capacitance as expected in a doped semiconductor \cite{Huang1992b,Hudlet1995}. 

In fact, the capacitance in a metal-insulator-semiconductor (MIS) structure --- physically represented by a
metal AFM tip probing a semiconductor sample --- exhibits three distinct regimes: accumulation, depletion
and inversion. Consider  a n-doped sample interacting with a positively biased tip: the mobile electrons
in the sample are attracted to the sample surface, resulting in an accumulation layer of the majority
carriers (electrons, in this case). Under negative tip bias, the majority carriers are repelled from the
sample surface, leading to a depletion layer. An increasingly negative bias leads to a thicker depletion
layer until the minority carrier density at the surface surpasses the initial majority carrier density, as a result of increasing band bending,
leading to an inversion layer \cite{Nicollian2002,Schwarz2000}. A similar understanding applies to
p-doped semiconductors. 

As such, to better probe the non-parabolic response in Figure \ref{fig:asym_bias} a), the numerically differentiated frequency shift with respect to the applied tip bias is plotted. The derivative is not linear over a -6~V to 8~V bias range. Instead, the derivative is comprised, piece-wise, of three lines with distinct slopes. 

\begin{figure}
\centering
\includegraphics[width=\linewidth]{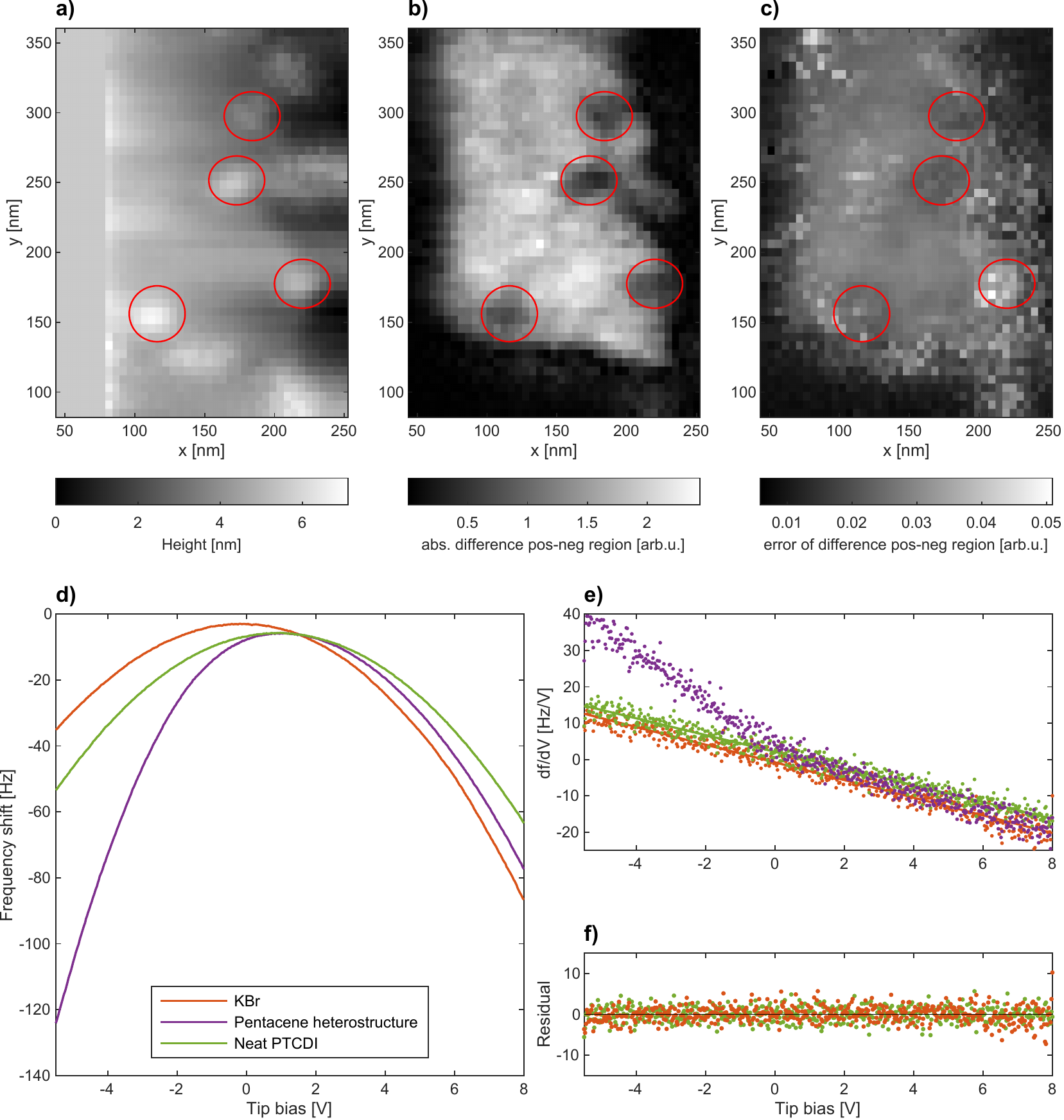}
\caption{a) Topography extracted from a KPFS map of pentacene/PTCDI island (bottom part of fig. \ref{fig:topo} a)). b) The difference in the derivative of the frequency shift for the positive and negative bias side clearly shows a difference for the pentacene island while no significant difference is observed for the KBr crystal and the PTCDI islands (red circle) connected to the pentacene island. c) Error of the calculated difference. d) KPFS measurements on KBr, doped pentacene, and neat PTCDI. The derivative of the frequency shift (e) reveals linear behavior for KBr and neat PTCDI, 
indicating parabolic behavior of the frequency shift. The doped pentacene does not show a linear behavior. f) The residual to a linear fit for KBr and neat PTCDI (red and green) is shown at the 
bottom. All data recorded at 0~nm tip lift from $\Delta f = -6~\text{Hz}$ setpoint.}
\label{fig:three_spots}
\end{figure}

We use the model present by Hudlet et al.\cite{Hudlet1995} which includes the voltage dependent capacitance to calculate the electrostatic force between a metal tip and a doped semiconductor as shown in \ref{fig:asym_bias} b). We add a parabolic frequency shift to the calculated signal to account for the capacitive force between the body of the tip and the sample as well as any background effects due to the KBr substrate.   The asymmetric shape of the bias spectroscopy measurement on pentacene can be qualitatively reproduced, emphasizing the need to consider the voltage dependent capacitance for our system. Despite the fact that the band bending and carrier diffusion in our system is essentially 3 dimensional, the qualitative agreement between the measurement and the model calculation that assumes one dimensional (plane-plane) system is notable. The difference in the slope of the derivative is not fully reproduced with this simple model, which we attribute to different carrier recombination times, unknown doping concentration, tip-sample capacitance, and neglected substrate effects. Lower intrinsic carrier densities and frequency shift as a function of surface potential are shown in the SI (fig S.2 and S.3).

The spatial variation of the non-parabolic KPFS data is illustrated in figure \ref{fig:three_spots}. We take the absolute difference of the fitted derivative of the frequency shift ($d\Delta f/dV$) obtained for individual fits to the positive and negative voltage window of the KPFS data (details in SI). The pentacene island exhibits a strong difference, with some variation within, while the PTCDI islands connected to it (red circles) do not show a significant difference and behave more in line with the KBr substrate. This indicates the doped semiconductor response is limited to the pentacene island.
\begin{figure*}
\centering
\includegraphics{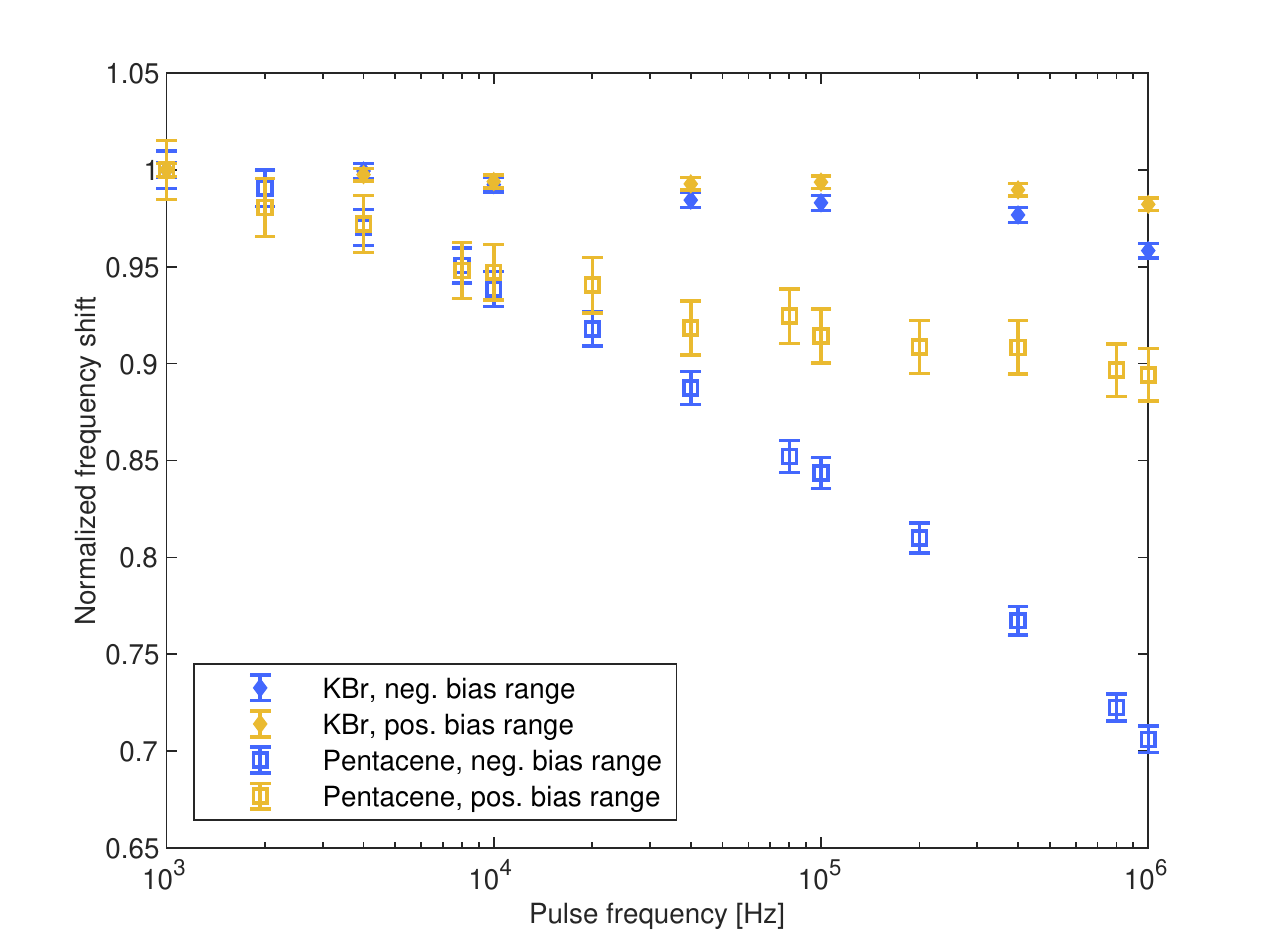}
\caption{A square wave voltage is applied to the tip and the frequency shift is recorded for a tested frequency range from 1~kHz to 1~MHz. The frequency shift is normalized to the lowest applied pulse 
frequency. Pentacene heterostructure exhibits a strong frequency response in the negative range (inversion) where as KBr shows no response. In the positive range (accumulation) a response 
is measured, reaching a plateau after 10 kHz.}
\label{fig:pulsed_bias}
\end{figure*}

It is well known that the capacitance of an MIS structure operating in the inversion regime exhibits a frequency dependence, whereas the capacitance in the accumulation-mode does not \cite{Nicollian2002,Wang2005}. In fact, pulsed MIS capacitance 
measurements have previously been used to determine the minority carrier lifetime in bulk samples 
\cite{Zerbst1966,Kano1972,Schroder1970,Schroder1972,Hillen1980,Ding1993}. To this end, we use a pulsed bias experiment to verify whether the sample is n-doped or p-doped. We implement this technique by holding the tip at a constant height (0~nm tip lift from $\Delta f = -6~\text{Hz}$ setpoint) above the pentacene island while a square wave bias pulse is applied to the tip and the frequency shift of the cantilever is recorded. The tip oscillation amplitude is set to 6~nm using a Nanosesors PPP-NCHPt cantilever with a resonance frequency of 288~kHz. We apply a 1~V to -8~V (negative) or a 1~V to 4~V (positive) amplitude square wave, while varying the frequency of the pulse (from 1~kHz to 1~MHz), to probe the negative/positive voltage range of figure \ref{fig:asym_bias} respectively.  A null experiment is also performed on KBr.

No significant change in frequency shift is observed for KBr for both positive and negative applied square waves, as expected. The pentacene/PTCDI heterostructure, on the other hand, exhibits a decrease in frequency shift with increasing pulse frequency for negative voltages. A decreased response for higher pulse frequencies is expected for inversion due to the finite life- and generation-time of the minority carriers needed to form the inversion layer. We can deduce that the inversion regime occurs for negative bias, which in turns corresponds to n-doped pentacene. As such, we can assign the three regimes identified in Figure~\ref{fig:asym_bias} a) to inversion (yellow), depletion (blue), and accumulation (green). The plateau reached at positive bias for the pentacene location can be attributed to the small contribution of minority carriers  due to the starting point (1V) of the pulsing being located in the depletion regime.

The exact lifetime of the minority carriers cannot be extracted from this measurement, since the exact doping concentration is unknown---this must be a known parameter for a quantitative result to be extracted from the pulsed MOS capacitance measurement presented by Zerbst \cite{Zerbst1966}. However, it can be concluded that the life- or generation-time of the minority carrier is in the order of tens of microseconds corresponding to a $1/e$ decay time in our measurements.

Furthermore, we must consider the subtle differences between the doped semiconductor measured here, and a MIS structure. For example, in contrast to a MIS field effect transistor (MOSFET), where the minority carriers are injected by the metal contact, no direct electrical connection between the doped semiconductor to the metal electrode exist in our system. The minority carriers needed to form the inversion layer are, in fact, supplied by the larger pentacene/PTCDI island. Only the area directly underneath our AFM tip is depleted of majority carriers. The time response in Figure \ref{fig:pulsed_bias} could therefore be limited by the minority carrier generation and diffusion in contrast to just the minority carrier lifetime.

\begin{figure*}
\includegraphics[width=1\linewidth]{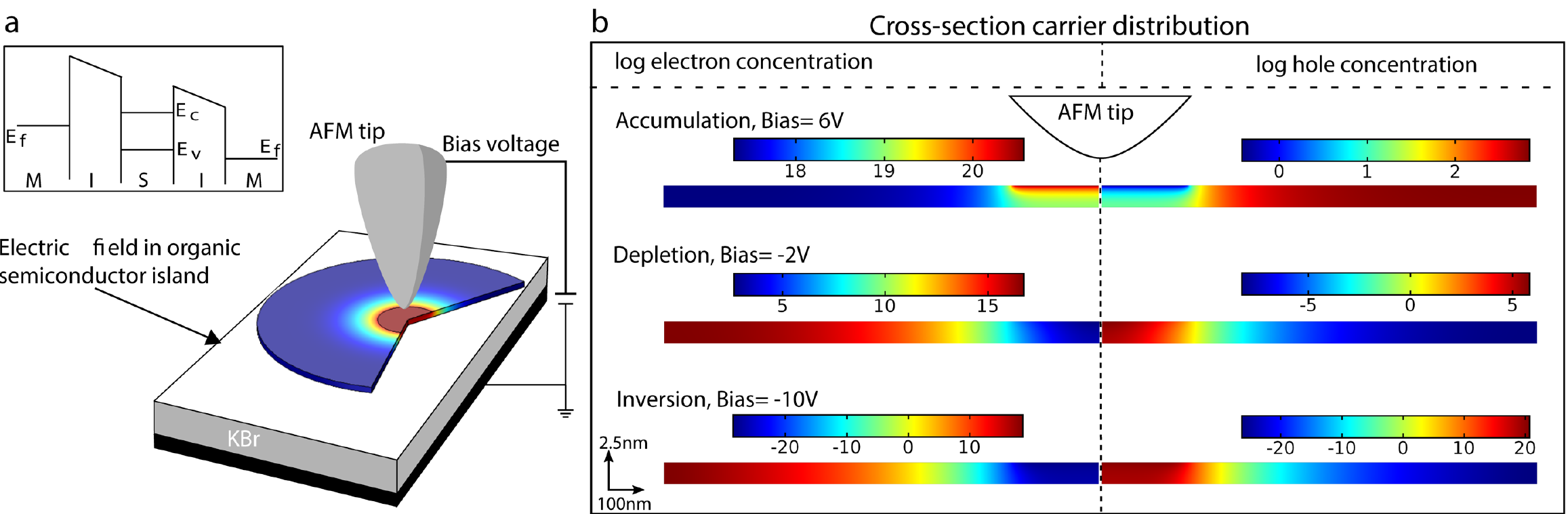}
\caption{a) A schematic of the experimental setup. The color-scale of the semiconductor corresponds to the electric potential in the sample at a bias voltage of -10V. The inset shows a band diagram of the sample structure with a organic semiconductor sandwiched between two insulators, followed by a metal contact.  b) Cross section of a finite element simulation, done using COMSOL, of the hole and electron concentration in a n-doped ($10^{16}\text{cm}^{-3}$), 2.5~nm thick Si sample, as shown on the left. Only the area underneath the tip is inverted when a positive bias is applied.}
\label{fig:simulation}
\end{figure*}

A schematic drawing of the experimental configuration of tip and sample can be seen in Figure~\ref{fig:simulation}. The corresponding band diagram, for a system consisting of a metal (AFM tip), insulator (vacuum), semiconductor (pentacene and PTCDI heterostructure), insulator (KBr substrate) and metal back electrode (sample holder) is also illustrated. As shown in Figure~\ref{fig:simulation}a, the electric field penetration range is confined to the area underneath the AFM tip. This implies the carriers density directly under the tip are exclusively altered by the incident electric field. For further illustration, we consider a simulation of the hole and electron concentration for a n-doped ($10^{16}$ cm$^{-3}$) 2.5~nm thick Si sample, on top of a 100~nm insulator ($\epsilon = 4.1$) supported by a metal electrode (0~V) with a top electrode (20~nm radius) separated by 3~nm of vacuum (Figure~\ref{fig:simulation}b). It is clear that only the area beneath the top electrode (tip) is inverted when a negative bias is applied. The hole concentration surpassed the initial dopant concentration at a bias of $-10$~V, therefore inverting the semiconductor underneath the tip to a p-type.

Theoretically, a non-parabolic response should only be observed over the doped semiconductor; measurements taken over neat PTCDI and KBr should reveal purely parabolic behavior as a function of bias, as indicated by Equation~\ref{eq:Felec}. Thus, we repeat the same KPFS measurements over clean KBr and neat PTCDI, and compare the results with data taken over the pentacene/PTCDI heterostructure (shown in Figure~\ref{fig:three_spots}). The respective measurement locations are indicated in Figure~\ref{fig:topo}a by three red crosses. As expected, we observe that frequency shift measurements taken over KBr and neat PTCDI are parabolic. Accordingly, the derivative of the frequency shift is purely linear over the applied bias range for the aforementioned data sets. In contrast, KPFS measurements taken over pentacene/PTCDI heterostructure show the same non-parabolic behavior as before. As a final test, the same experiment was repeated with different cantilevers (with a new Pt-Ir coated cantilever (PPP-NCHPt) and a silicon cantilever (PPP-NCHR)) for isolated, separate pentacene/PTCDI islands, all to reveal the same results. 
 While it might appear as if the n-doping of the pentacene layer is due to the PTCDI islands, we have found non-parabolic behaviour in neat pentacene islands as well. We attribute the origin of the non-parabolic behaviour to impurities found in the pentacene island. These sites appear as locations of charging, where charging rings can be observed in the dissipation signal of the AFM (see supplementary information). While the nature of these sites is of great interest and under investigation, this publication focused on the global effect of these sites on the pentacene layer.
 KPFS measurements on neat pentacene before the deposition of PTCDI follow a parabolic behavior (see supplementary material), as long as there are no defects site appearing as charging rings in the dissipation channel, confirming our conclusion of these charging sites acting as impurities and thereby n-doping the affected pentacene islands.

In conclusion, we measure the electrostatic force of an accumulation, depletion and inversion layer in a 2.1 nm thick organic semiconductor. Impurities in the pentacene island, showing up as charging rings in the dissipation channel, act as a dopant for the much larger pentacene island, resulting in a n-doped pentacene island. The inversion layer is generated by sweeping the bias voltage on an AFM tip, which is in close proximity to the island. Only the area directly underneath the AFM tip (approximate radius of 20~nm) is depleted of majority carriers. The measurement qualitatively agrees with numerical simulations. Exact numerical calculations are not possible for our system, since various parameter such as the dopant concentration, intrinsic carrier density, KBr capacitance contribution, etc. are unkown. 

The formation of the inversion layer was further verified by applying a pulsed voltage and measuring the frequency response. 
Observing the movement of carriers in organic semiconductor and the formation of inversion layer at this length scale is a crucial step towards studying organic photovoltaics at the nanometer scale. Our study shows that a 2.1 nm thick n-doped pentacene layer can locally be brought to inversion at an approximate 20~nm lateral size. This would enable the production of an organic FET with a channel  on the same length scale, which has thus far not been achieved.
Further investigation of this system under light illumination is of great interest to advance the understanding of the local nanometer behavior of organic semiconductor during charge generation. Experiments conducted under illumination could potentially help to determine the minority carrier generation- and lifetime of organic semiconductors at the nanometer scale.

\section*{Acknowledgments}

The authors would like to thank NSERC and FQRNT for funding. Z.S. gratefully acknowledges the SNSF for a DOC.Mobility fellowship. 

\bibliography{library}

\clearpage
\newpage
\renewcommand\thefigure{S.\arabic{figure}}    
\setcounter{figure}{0}
\widetext
\section*{Supplementary information}

\subsection*{Difference map of $\frac{df}{dV}$}

A bias spectroscopy is recorded at each point of the map. The negative ($-5.2622~V$ to $-1.5636~V$) and positive ($2.3992~V$ to $6.8904~V$) side of the KPFS curve are fitted independently to equation 1. For the map in figure 3, the absolute difference between the fitted value for $\frac{d^2C}{dz^2}$ for positive and negative side is calculated, along with the error from the fit confidence interval (95\%).

\subsection*{Calculation of electrostatic force and frequency shift}

The electrostatic force was calculated following the one-dimensional model of MIS structure 
presented by Hudlet et al. \cite{Hudlet1995}. 
The force acting on the tip is expressed by 

\begin{equation}
    F_{elec} = - \frac{Q_s^2}{2\epsilon_0}
\end{equation}
where $Q_s$ is the total charge in the semiconductor and given by
\begin{equation}
    Q_s = - \text{sgn}(u)\frac{kT}{q}\frac{\epsilon}{L_D} 
    \left( e^u - u -1 + \frac{n_i^2}{N_D^2}(e^{-u} + u -1) \right)^{1/2}.
\end{equation}
Here, $u=(qV_s/kT)$ is the reduced surface potential, 
$L_D = (\epsilon kT/2N_D q^2)^{1/2}$  Debye length, 
$N_D$ the dopant density, 
$q$ is the elementary charge.

For each applied bias voltage $V_0$,  a surface potential $V_s$ is calculated 
by numerically solving the following equation    
\begin{equation}
    V_0 = V_s - \frac{Q_s(V_s)}{C_l}
\end{equation}
for a given tip sample distance. 
Here $C_l$ is the air gap capacitance per unit area in series with the capacitance of the alkali halide substrate.
The calculation is performed for a range of bias voltage with a range of tip-sample distance
to obtain a bias vs force curve. 
The calculation is repeated for a range of tip-sample distance, $z$ to calculate the frequency shift.

The present calculation in figure 2 b) of the main text, was shifted by 1.5V to account for the CPD measured in our experiments. Additionally we added a parabolic background signal with a CPD of -0.5V to account for any capacitive force between the body of the tip and the sample as well as any background effects due to the large KBr substrate. The dopant concentration of $5*10^{16}~cm^{-3}$ was used assuming 5 charging rings (1 ring = 1 dopant) and a volume of $10^{-16}~cm^3=400~nm\times120~nm\times2.1~nm$ for the pentacene island.

The calculation of the frequency shift versus bias voltage curves was then repeated 
for different dopant densities to observe the evolution 
from a symmetric to non-symmetric bias curve. 
Below a selection of the performed calculation is shown to illustrate this transition.
An intrinsic carrier density of $1\times 10^{10}$/cm$^3$ was used for all calculation. 
The bias curves evolve from a almost symmetrical shape to an asymmetrical shape 
while the dopant density changes from $1$ to $100$. 
The calculations with a dopant density of $1\times 10^{15}$/cm$^3$ was used in the main text.

\begin{figure}
\centering
\includegraphics[width=1\linewidth]{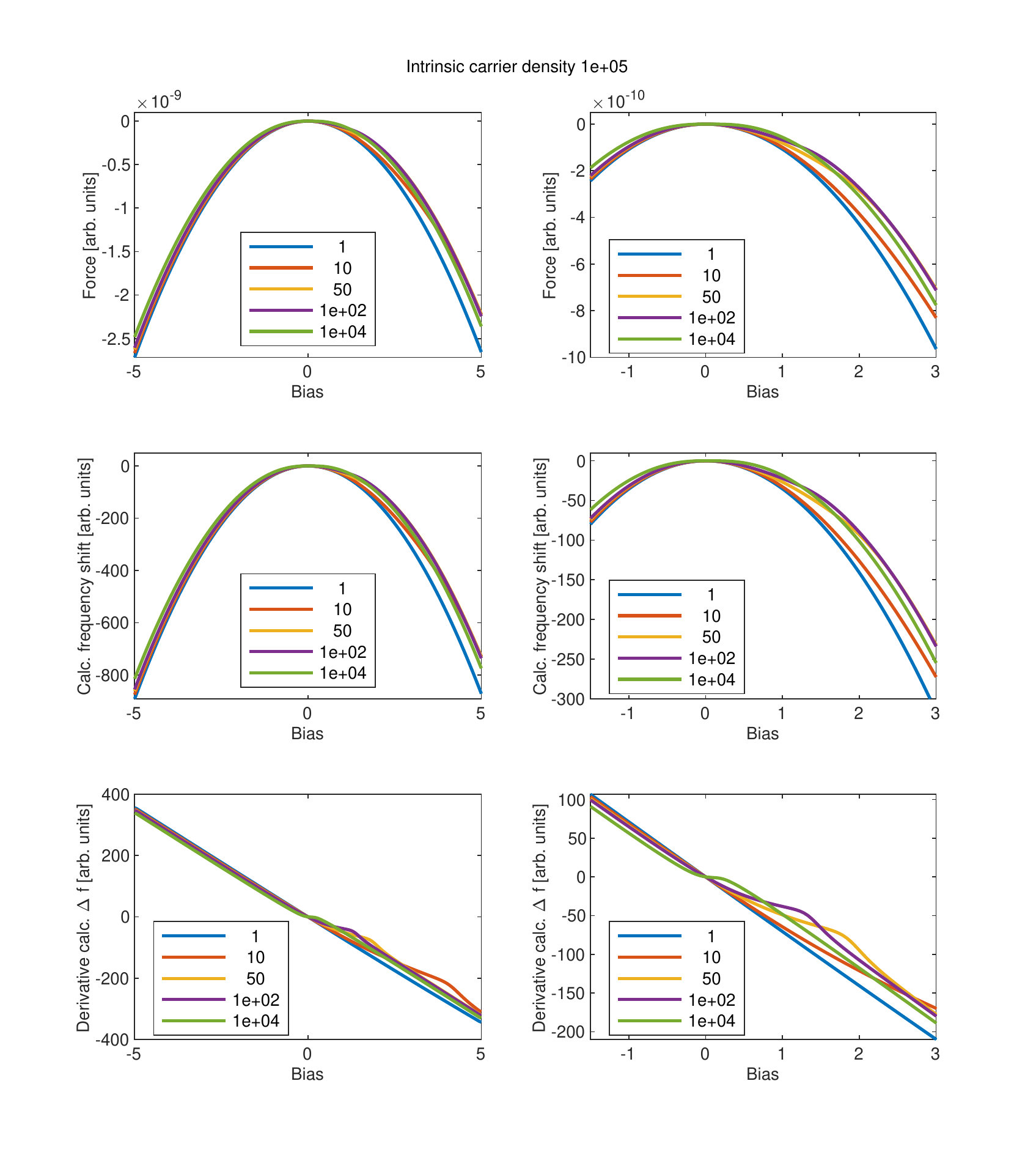}
\caption{Calculated frequency shift vs bias curves for different dopant densities indicated in the legend at a lower intrinsic carrier density of $10^5$. Top shows force vs bias, middle frequency shift vs  bias. Left column shows a smaller bias range for better illustration of the change. The bottom shows the derivative of the calculated curves above.}
\label{fig:BiasSpec_calc1_sum}
\end{figure}

\begin{figure}
\centering
\includegraphics[width=1\linewidth]{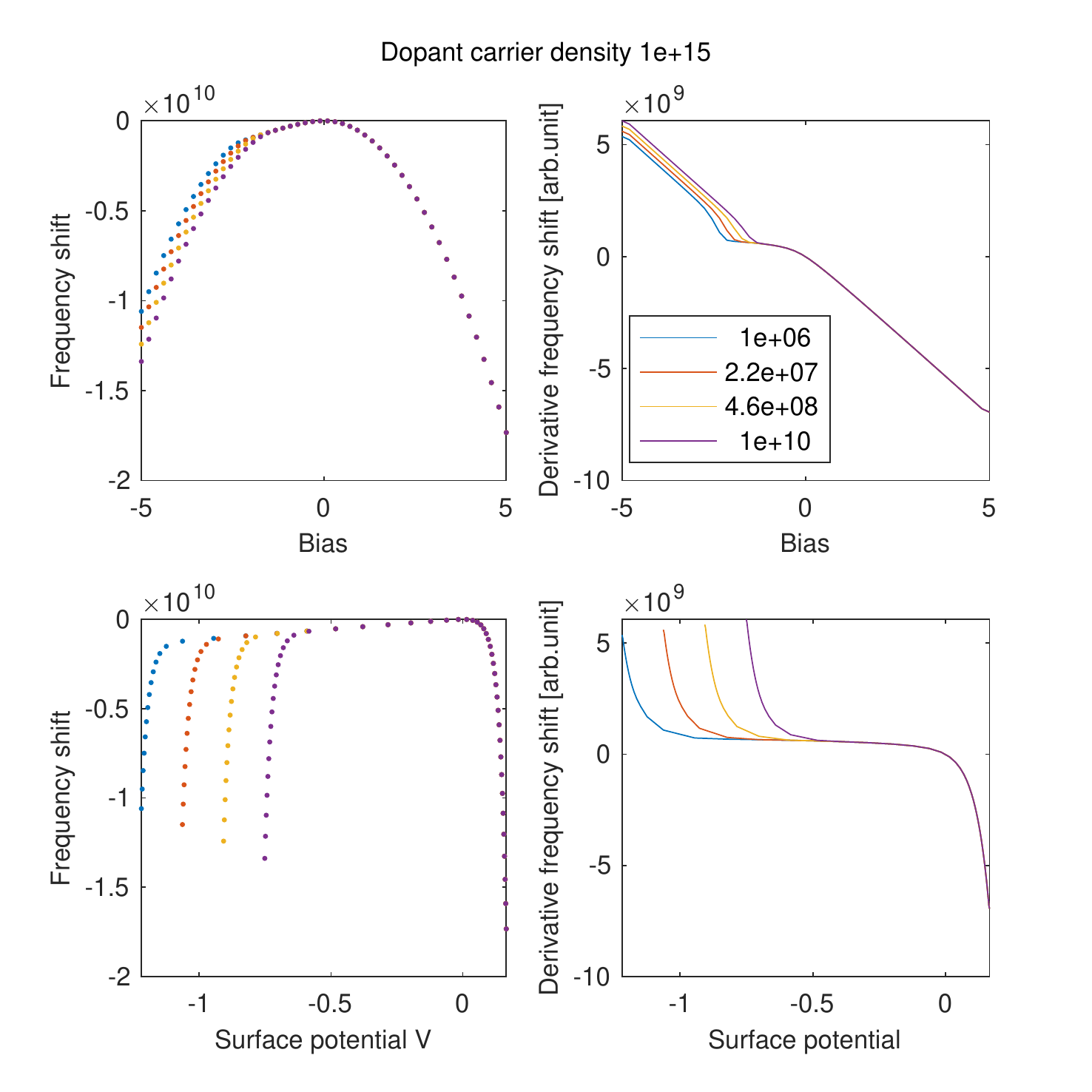}
\caption{Calculated frequency shift for different dopant densities indicated in the legend. Top row shows frequency shift vs bias, bottom row shows the same calucaltion plotted vs surface potential.}
\label{fig:SI_highND}
\end{figure}

\begin{figure}
\centering
\includegraphics[width=1\linewidth]{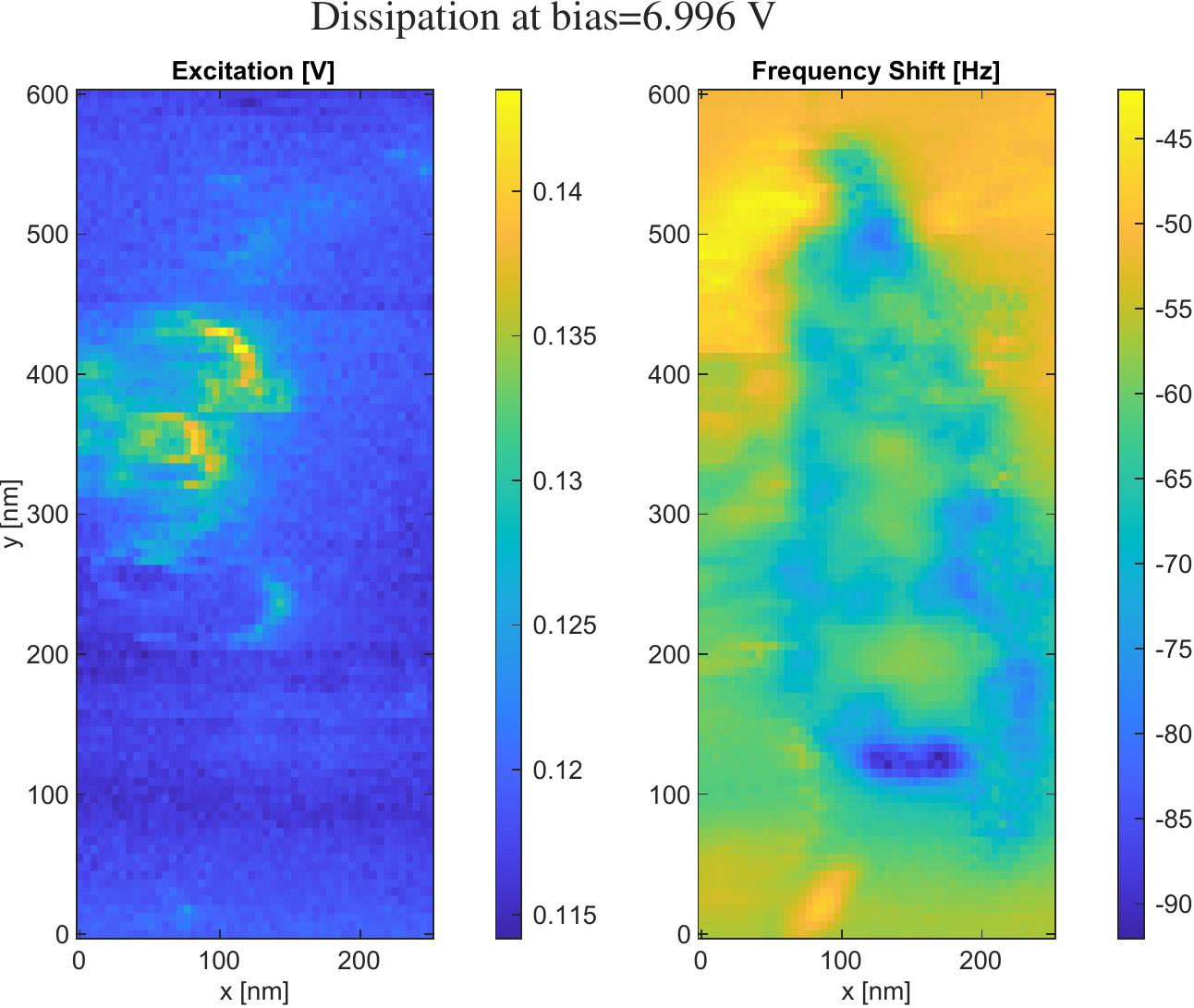}
\caption{Excitation (left) and frequency shift(right) extracted from a bias spectroscopy map on the pentacene/PDTCI heterostructure shown in figure 1a). A multitude of charging rings \cite{miyahara_quantum_2017} can be seen in the excitation channel indicating impurities in the pentacene layer. Oscillation amplitude 6~nm, frequency shift setpoint -6Hz, tip lift 0~nm, applied bias 6.996~V}
\label{fig:charging_rings}
\end{figure}

\end{document}